\documentclass[10pt,conference]{IEEEtran}
\usepackage[T1]{fontenc}
\usepackage{cite}
\usepackage{caption}
\usepackage{algcompatible}
\usepackage{mathtools}
\usepackage{stackengine}

\usepackage{amsmath,amssymb,amsfonts}
\usepackage{amsmath,epsfig,cite,amsfonts,amssymb,psfrag,subfigure}
\usepackage{graphicx}
\usepackage{blindtext}
\usepackage{textcomp}
\usepackage[dvipsnames]{xcolor}
\usepackage{algorithm}
\usepackage{algorithmicx}
\usepackage[noend]{algpseudocode}
\usepackage{amsthm}
\def\BibTeX{{\rm B\kern-.05em{\sc i\kern-.025em b}\kern-.08em
    T\kern-.1667em\lower.7ex\hbox{E}\kern-.125emX}}
\usepackage{balance}
\allowdisplaybreaks

\usepackage{booktabs,siunitx}

\usepackage{xcolor}
\def\BibTeX{{\rm B\kern-.05em{\sc i\kern-.025em b}\kern-.08em
    T\kern-.1667em\lower.7ex\hbox{E}\kern-.125emX}}
\usepackage[skip=1pt]{caption}
\begin{document}

\title{Moving Target Defense based Secured Network Slicing System in the O-RAN Architecture\vspace{-.1cm}
}
\author{\small Mojdeh Karbalaee Motalleb$^{\dagger}$, Chafika Benza\"{i}d$^*$, Tarik Taleb$^*$,  Vahid Shah-Mansouri$^\dagger$ \\ Email:  \{mojdeh.karbalaee,vmansouri\}@ut.ac.ir, \{chafika.benzaid, tarik.taleb\}@oulu.fi \\
\IEEEauthorblockA{ $^\dagger$School of ECE, University of Tehran, Tehran, Iran\\ $^*$University of Oulu, Oulu, Finland}
\vspace{-1cm}
} 
\maketitle
\pagenumbering{gobble}
%\vspace*{-2em}
\begin{abstract}
%\vspace*{-1.5em}
The open radio access network (O-RAN) architecture's native virtualization and embedded intelligence facilitate RAN slicing and enable comprehensive end-to-end services in post-5G networks. However, any vulnerabilities could harm security. Therefore, artificial intelligence (AI) and machine learning (ML) security threats can even threaten O-RAN benefits. This paper proposes a novel approach to estimating the optimal number of predefined VNFs for each slice while addressing secure AI/ML methods for dynamic service admission control and power minimization in the O-RAN architecture. We solve this problem on two-time scales using mathematical methods for determining the predefined number of VNFs on a large time scale and the proximal policy optimization (PPO), a Deep Reinforcement Learning algorithm,  for solving dynamic service admission control and power minimization for different slices on a small-time scale. To secure the ML system for O-RAN, we implement a moving target defense (MTD) strategy to prevent poisoning attacks by adding uncertainty to the system. Our experimental results show that the proposed PPO-based service admission control approach achieves an admission rate above 80\% and that the MTD strategy effectively strengthens the robustness of the PPO method against adversarial attacks.
%Three scenarios are considered: a system without an attack, an insecure system with an attack, and an MTD-secured system with an attack. The performance of an unsecured system drops dramatically after an attack. The numerical results indicate significant improvements after using novel MTD techniques.
 
\end{abstract}
 \vspace{-.3cm}
\section{Introduction}
A sixth generation (6G) wireless network will offer enhanced network capacity of $10\text{Gbps/m}^{3}$, lower end-to-end latency below $1$  ms, and increased data rates up to $1$ Tbps. The 6G capabilities will unlock new applications and services, including holographic communications, wireless brain-machine interaction,  autonomous driving, etc.~\cite{slawomir20216g}. 

6G networks will use network slicing to meet the varying QoS requirements of envisioned applications/services by dynamically creating logically isolated, service-tailored virtual networks (i.e., slices) on shared physical infrastructures~\cite{wu2022ai}.  A network slice instance consists of chained network functions and the required resources (e.g., compute, bandwidth, storage), spanning multiple technology domains (e.g., radio access network (RAN), core network (CN), and transport network). Despite its maturity in CN, network slicing remains challenging in other domains. The native virtualization and embedded intelligence of the open RAN (O-RAN) architecture are vital features to promote RAN slicing, enabling the delivery of genuinely end-to-end services to become a reality~\cite{oranMe, benzaid2022ai,oran2022slice}. Specifically, O-RAN architecture introduces RAN Intelligent Controller (RIC). This software-defined network controller leverages the capabilities of Artificial Intelligence (AI) and Machine Learning (ML) to enable intelligent and closed-loop RAN resource management and optimization.
The RIC is divided into non-real-time (Non-RT) RIC and near-real-time (Near-RT) RIC, which incorporate rApps and xApps, custom micro-service-based applications, operating on Non-RT scale ($>1s$) and Near-RT scale ($10 - 1000ms$), respectively. 
% The RIC brings flexibility, adaptability, and automation capabilities required to empower dynamic and real-time control of RAN resources to meet the differentiated service level agreement (SLA) on different slices.

%<Now we should introduce the problem of slice admission and emphasize the role of DRL and its integration in RIC>
% \subsection{\textcolor{blue}{Related Literature}}
% \subsubsection{\textcolor{blue}{RAN Slicing and VNF placement}}

In RAN slicing, efficient slice-aware resource management through slice admission control is crucial. 
% \textcolor{blue}{The optimization method and game theory are still techniques for solving these problems.}
Recently, the potential of ML and, more particularly, Deep Reinforcement Learning (DRL) techniques have been explored for enabling optimal slice admission control strategies in the O-RAN system. 
% A deep transfer RL is assumed in paper \cite{zhou2022learning} for joint radio and cache resource allocation in RAN slicing.
A federated DRL is presented in paper \cite{zhang2022federated} to manage multiple independent xApps in O-RAN for network slicing. Two xApps, which are jointly communicated, are implemented for power and physical resource block management. 
In \cite{oranMe}, the problem of obtaining the optimal number of virtual network functions (VNFs) and baseband resource management is considered for the RAN slicing in the O-RAN architecture, which is solved using an optimization technique. 
% In the paper \cite{noroozi2019service}, the problem of service admission control is assumed for the RAN and CN using the knapsack algorithm.
% In \cite{wang2017online}, the problem of the online VNF placement and service function chaining is considered to minimize the prediction errors in the VNF demands.
% The problem of the routing system is also solved in this paper using an online learning technique.
In \cite{luu2022admission}, an optimization technique for the infrastructure resource reservation is used to prioritize admission control for multiple slices. %The authors introduce a game theory network-slicing framework in \cite{caballero2018network}, including admission control and resource allocation.}
% \subsubsection{\textcolor{blue}{Security Issues and MTD technique}}

ML techniques, including DRL, face vulnerabilities to adversarial attacks that manipulate data during (re)training or serving~\cite{aisecme, zsmme1, zsmme2}. For example, resource manipulation could mislead a DRL-based slice admission model, wrongly rejecting RAN slice requests. Ensuring ML security is crucial for O-RAN integration, building trust in their decisions
%\textcolor{red}{
Addressing O-RAN's security from an ML perspective involves multiple methods: Zero Trust (ZT), blockchain, and Moving Target Defense (MTD).
ZT adopts a proactive 'never trust, always verify' stance to fortify O-RAN infrastructure against ML threats.
Blockchain ensures data integrity and transparent model history, enhancing trust and accountability in ML processes.%\cite{blockchain1}.%}
An effective defense strategy discussed in~\cite{aisecme} is the Moving Target Defense (MTD) paradigm. MTD enhances security by continually changing the attack surface, making it harder for attackers to predict and exploit vulnerabilities. This approach has gained in safeguarding ML models, particularly in computer vision and malware domains(e.g.~\cite{sengupta2019mtdeep, rashid2022mtd} ).

%\textcolor{red}{
The MTD technique provides a dynamic and proactive security approach, distinguishing it from ZT and blockchain, which primarily concentrate on access control and data integrity. MTD's continuous alteration of the attack surface poses a formidable challenge for adversaries, enhancing resilience against evolving threats. Consequently, our paper centers on employing MTD to secure the O-RAN system.%}
% To build robust attack graphs in an SDN environment, the authors presented a scalable port-hopping MTD using vulnerability scores \cite{chowdhary2016sdn}.
In \cite{qiu2021mt}, the authors consider a Trojaning attack defense framework based on an MTD on the Deep neural network (DNN), which randomly selected dimensions in multidimensional training models. According to the results, they guarantee DNN's availability and protect it from Trojan attacks.
%\subsection{\textcolor{blue}{Motivations}}

Currently, previous works focus on traditional RAN slicing with classic methods such as optimization and game theory or artificial intelligent/Machine learning (AI/ML) techniques.
Although some of them highlight O-RAN slicing, they do not mention the softwarization of the O-RAN architecture in RAN slicing technology. Moreover, the lifecycle of slicing is not assumed in previous works.
In contrast, our paper differs from current papers by focusing on the planning and creation phase of RAN slicing to estimate the optimal number of predefined VNFs in the O-RAN distributed unit (O-DU) and the central unit (O-CU) for different slices based on the processing delay threshold of the system, which is done on a large time scale. Afterward, we consider the managing phase of the RAN slicing lifecycle by analyzing the dynamic service admission control and power minimization on a small time scale for different services with different QoS in the processing layer of the O-RAN technology using the DRL technique. 
Due to the dynamic nature of the problem, the sequential decision-making, and the large state space, DRL is the most appropriate approach.
Moreover, to the best of our knowledge, none of the existing contributions has considered the security issues stemming from using DRL techniques in the RAN slicing.
The adversarial defense strategy employs MTD to bolster the proactive resilience of our DRL model against attacks. Implementing MTD entails training various models with similar performance for different xApps. The xApps will be randomly selected after learning. The main contributions of this paper are as follows:
\vspace{-.1cm}
\begin{itemize}
\item We study the problem of estimating the optimal number of VNFs in each slice and solving the secure dynamic service admission control and power consumption in the O-RAN using the RAN slicing.
\item The problem of estimating the optimal pre-defined VNFs are solved mathematically on a large time scale to obtain how many VNF chains can be deployed for each slice. In contrast, the problem of service admission control and minimizing total power is solved dynamically on a small time scale using the PPO algorithm, an actor-critical DRL technique.  
\item We introduce a novel defense strategy that relies on the MTD paradigm to make the proposed DRL approach resilient to adversarial attacks to prevent degrading system utilization. Rather than shuffling the network as done in prior MTD studies, we shuffle the AI models for frequent system changes. The developed MTD strategy consists in dynamically picking a model from a set of PPO models trained with different configurations, increasing the adversary's uncertainty.
\item The numerical results demonstrate the improvement in the PPO method against the baseline method and the negative impact of adversarial attacks on a PPO-based service admission control system without appropriate defenses. They also show the effectiveness of the novel MTD-based defense strategy in enhancing the solution's robustness against those attacks.
\end{itemize}

The rest of the paper is organized as follows. Section~\ref{system} introduces the system model. Section~\ref{drlsac} formulates the target service admission control problem and presents the proposed DRL approach. Section~\ref{secur} describes the adversarial attack model and the devised MTD-based defense strategy. Section~\ref{numerical} and \ref{conclud} discuss the numerical results and conclusion.    
%and problem formulation are discussed in Section~\ref{system} and Section~\ref{probfor}. The proposed method is introduced in Section~\ref{proposed}.
%The details of the attacked system and secured method is described in Section~\ref{secur}.
%Finally, the numerical results and the conclusion are discussed in Section~\ref{numerical} and Section~\ref{conclud}, respectively.

% vnf placement +
% deep rl +
% security risk oran +
% MTD method and others +
% main contribution
\begin{figure}
  \centering
  \captionsetup{justification=centering}
    \includegraphics[scale = .22]{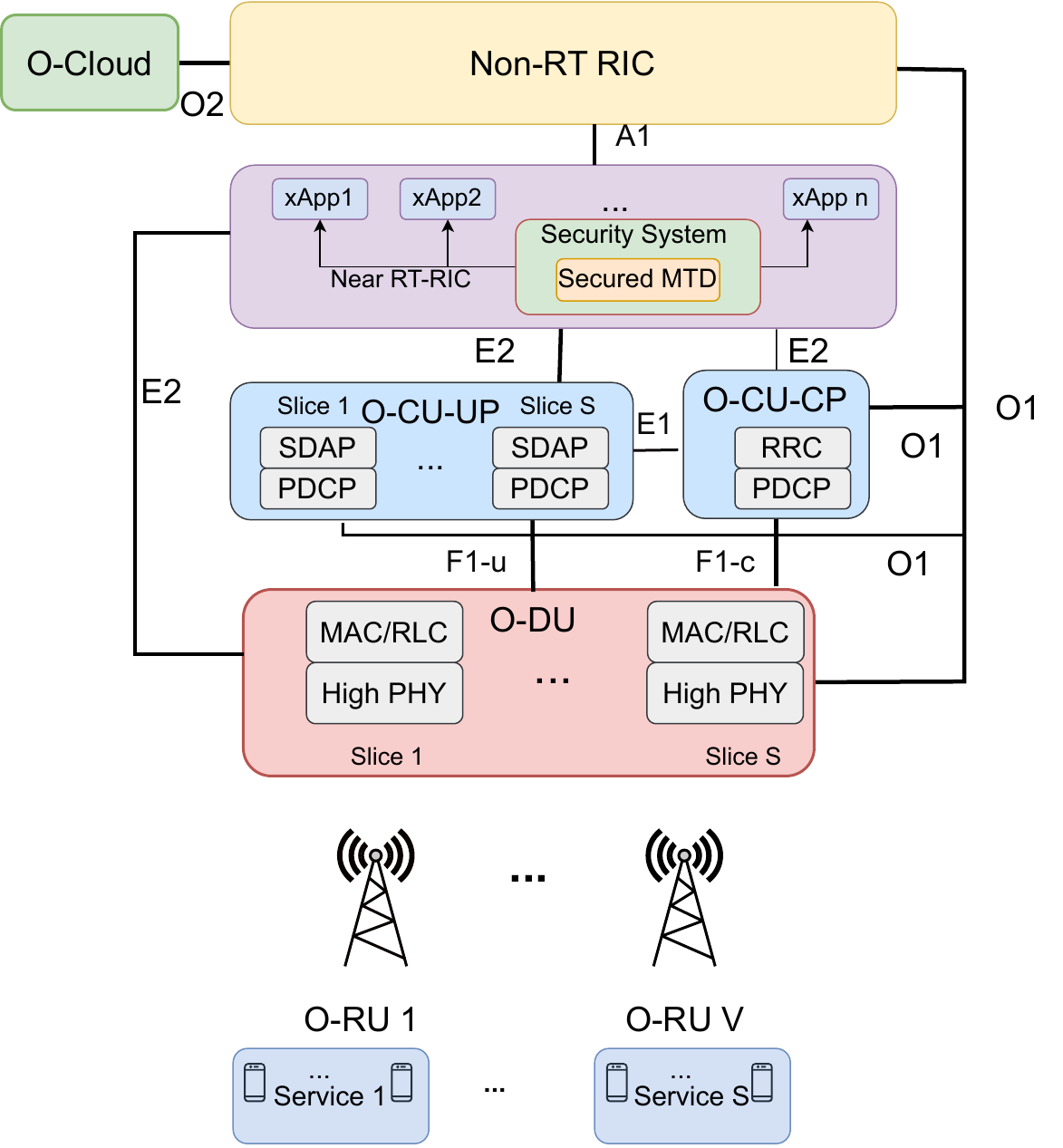}
  \caption{The secured intelligent O-RAN architecture.}
  \label{fig:c15}
  \vspace{-0.4cm}
\end{figure}
\vspace{-.2cm}
\section{System model}\label{system}
As illustrated in Fig. \ref{fig:c15}, we consider the dynamic slice admission control and power optimization in the O-RAN system. Assume there are $S$ pre-defined RAN slices serving $S$ services. Each instantiated RAN slice comprises several VNFs providing the services of the different virtualized O-RAN units, such as the O-DU and O-CU. Note that O-DU runs the high Physical (PHY), Medium Access Control (MAC), and Radio Link Control (RLC) layers. To deliver different services, MAC and RLC are deployed on isolated VNFs. The O-DU provides services to users through the radio unit (O-RU), which contains low PHY and radio frequency. The O-CU contains the O-CU control plane (O-CU-CP) and the O-CU user plane (O-CU-UP), which handle the control and data messages, respectively. The O-CU-CP includes packet data convergence protocol (PDCP) and radio resource control (RRC), and the O-CU-UP contains PDCP and service data adaptation protocol (SDAP) which are deployed on isolated VNFs for different services. As shown in Fig. \ref{fig:c15}, the virtualized O-RAN functions can be dedicated to each slice (e.g., O-DU, O-CU-UP) or shared between slices (e.g., O-CU-CP).

%In this section, we want to analyze the problem of secured dynamic service admission control and VNF placement in the O-RAN architecture. 
%Here, we describe the system model of the network slicing O-RAN architecture, as shown in Fig. \ref{fig:c15}.
% As a final step, secure techniques are presented in order to make the system safe from vulnerabilities.
%\subsection{System Model}

%Assume there are $S$ pre-defined slices serving $S$ services.  
%The O-RAN architecture comprises two processing layers as logical nodes: a centralized unit (O-CU) and a distributed unit (O-DU). 

%The O-DU contains the medium access control (MAC), radio link control (RLC), and high physical (high PHY) layer. To provide different services, MAC and RLC are deployed on isolated VNFs.

%Furthermore, the O-DU provides services to users through the radio unit (O-RU) which contains low PHY and radio frequency (RF). 

%The O-CU contains the O-CU control plane (O-CU-CP) and the O-CU user plane (O-CU-UP), which handle control and data messages. The O-CU-CP includes packet data convergence protocol (PDCP) and radio resource control (RRC), and the O-CU-UP contains PDCP and service data adaptation protocol (SDAP) which are deployed on isolated VNFs for different services.

The O-RAN system uses Near-RT RIC and Non-RT RIC to control the O-DU and O-CU for resource management. The Near-RT RIC hosts third-party applications xApps to provide management and optimization services. Here, we consider the deployment of xApps providing DRL-based resource management services. The Non-RT RIC offers offline ML models to support Near-RT RIC functions.  

%The O-RAN architecture uses near-real-time (near-RT) and non-real-time (non-RT) RICs to control the O-DU and the O-CU for resource management.
%An x-application (xAPP) is an application that provides network management and optimization services by third-party vendors which is deployed in the near-RT RIC. 
%The xAPPs provide resource management methods based on RL.
%In addition to near-RT RIC functions, non-RT RICs offer assistance and ML models to support near-RT RIC \cite{zhang2022federated, zhang2022team}.
We assume different services that use isolated pre-defined slices in this system model.
The pre-defined slices contain reserved VNFs for two logical nodes of MAC/RLC functions in the O-DU and PDCP/SDAP functions in the O-CU-UP.
% \begin{itemize}
%     \item MAC/RLC functions in the O-DU
%     \item PDCP/SDAP functions in the O-CU-UP
% \end{itemize}
We consider a simple service function chain in the O-DU and O-CU.
Suppose we have $M_s^d$ and $M_s^c$ VNFs in the O-DU and O-CU-UP processing layer for the service $s$.
%\vspace{-.5cm}
\subsection{Mean Delay}
\vspace{-.0cm}
Consider the mean arrival rate of the service $s$ is Poisson with rate $\bar{\alpha}_s^c$ in the O-CU-UP layer. The mean arrival data rate of the $s^{th}$ service in the O-DU is approximately equal to the mean arrival data rate of the $s^{th}$ service in the O-CU-UP ($\bar{\alpha} = \bar{\alpha}_s^d \approx \bar{\alpha}_s^c$). This is because the amount of data transmitted through the route (despite frame changes ) is constant. 

Incoming traffic to VNFs is divided equally by load balancers at each layer for each service. Assume that each VNF's baseband processing is represented by an M/M/1 queue. In each slice, one VNF processes each packet. Accordingly, the mean delay for slice $s$large time scale in the O-DU and O-CU can be calculated as M/M/1 queue \cite{oranMe} as follows
as $\bar{T}_{s}^{DU} = \frac{1}{\bar{\mu}_s^d - \bar{\alpha}_{s}/{M_s^{d}}},$ and $\bar{T}_{s}^{CU} = \frac{1}{\bar{\mu}_s^c - \bar{\alpha}_{s}/{M_s^{c}}}.$
% \begin{equation}
% \begin{split}
% \bar{T}_{s}^{DU} &= \frac{1}{\bar{\mu}_s^d - \bar{\alpha}_{s}/{M_s^{d}}},\\
% \bar{T}_{s}^{CU} &= \frac{1}{\bar{\mu}_s^c - \bar{\alpha}_{s}/{M_s^{c}}}.
% \end{split}
% \end{equation}
In addition, $\frac{1}{\bar{\mu}_s^d}$ and $\frac{1}{\bar{\mu}_s^c}$ are the mean service time of the system in the O-DU and the O-CU-UP layer in the lrge time scale, respectively. 
For the simplicity, we assume that the O-CU and the O-DU have the same processing system. Hence $\frac{1}{\bar{\mu}_s^c}\approx \frac{1}{\bar{\mu}_s^d}$. 
Therefore, we can consider that the $M_s = M_s^{d} = M_s^{c}$, for the simplicity.
As a result, $\bar{T}_s = \bar{T}_{s}^{DU} =\bar{T}_{s}^{CU}$.
Consequently, the mean total delay of the system in the slice s is $\bar{T}_s^{tot} = 2\times \bar{T}_s$. 
\vspace{-.1cm}
\subsection{Physical Data Center Resources}
%\vspace{-.1cm}
The VNF instances are also hosted on VMs that use data center resources.
Each VNF in each layer requires specific physical rescurces including CPU, storage, and memory based on the service requirements.
Consider a set of a tuple that expresses the instant required resources for VNF $m$ in the service function chain of the $\mathfrak{z} \in \{c,d\}$ (VNFs of O-DU or O-CU) in slice $s$ as $    \bar{\psi}_{s}^{m^\mathfrak{z}} = \{\psi_{\mathsf{C},{s}}^{m^\mathfrak{z}}, \psi_{\mathsf{S},{s}}^{m^\mathfrak{z}}, \psi_{\mathsf{M},{s}}^{m^\mathfrak{z}} \}.$
% \begin{equation}
%     \bar{\psi}_{s}^{m^\mathfrak{z}} = \{\psi_{\mathsf{C},{s}}^{m^\mathfrak{z}}, \psi_{\mathsf{S},{s}}^{m^\mathfrak{z}}, \psi_{\mathsf{M},{s}}^{m^\mathfrak{z}} \},
% \end{equation}
where, $\psi_{\mathsf{C},{s}}^m$, $\psi_{\mathsf{S},{s}}^m$, and $\psi_{\mathsf{M},{s}}^m$, provide the amount of CPU, storage, and memory that are required for the VNFs of the O-DU or O-CU ($\mathfrak{z} \in \{c,d\}$) .
Moreover, $\bar{\psi}_{s}^m \in \mathbb{C}^{3}$.
Accordingly, we indicate the total amount of CPU, storage, and memory, respectively ($\mathfrak{h} \in \{C, S, M\}$), for the O-DU and O-CU layers ($\mathfrak{z} \in \{c, d\}$) as $\bar{\psi}_{{\mathfrak{h}},s}^{\mathfrak{z},tot} = \sum_{m=1}^{M_s^\mathfrak{z}}{\psi}_{{\mathfrak{h}},s}^{m^\mathfrak{z}}, \;\; \mathfrak{z} \in \{c, d\}, \mathfrak{h} \in \{C, S, M\}$.
%\vspace{-.2cm}
% \begin{equation}
%     \bar{\psi}_{{\mathfrak{h}},s}^{\mathfrak{z},tot} = \sum_{m=1}^{M_s^\mathfrak{z}}{\psi}_{{\mathfrak{h}},s}^{m^\mathfrak{z}}, \;\; \mathfrak{z} \in \{c, d\}, \mathfrak{h} \in \{C, S, M\}
% \end{equation}
% \begin{subequations}
% \begin{alignat}{4}
%    \bar{\psi}_{{\mathsf{C}},s}^{\mathfrak{z},tot}& = \sum_{m=1}^{M_s^\mathfrak{z}}{\psi}_{{\mathsf{C}},s}^{m^\mathfrak{z}}, \;\; \mathfrak{z} \in \{c, d\} \\
%    \bar{\psi}_{{\mathsf{S}},s}^{\mathfrak{z},tot}& = \sum_{m=1}^{M_s^\mathfrak{z}}{\psi}_{{\mathsf{S}},s}^{m^\mathfrak{z}}, \;\; \mathfrak{z} \in \{c, d\}\\
%    \bar{\psi}_{{\mathsf{M}},s}^{\mathfrak{z},tot}& = \sum_{m=1}^{M_s^\mathfrak{z}}{\psi}_{{\mathsf{M}},s}^{m^\mathfrak{z}} \;\; \mathfrak{z} \in \{c, d\}, 
% \end{alignat}
% \end{subequations}
% where, $\mathfrak{z} \in \{c, d\}$ represents the processing layer of O-DU and O-CU.
% $\mathfrak{h} \in \{C, S, M\}$ shows the resource of CPU, Memory and storage.

Suppose we have $N$ data centers for the VNFs of the O-DU and the O-CU.
Each data center $n$, has a set of a tuple that expresses the amount of CPU, storage, and memory resources as ${\chi}_{s}^n = \{\chi_{\mathsf{C},{s}}^n, \chi_{\mathsf{S},{s}}^n, \chi_{\mathsf{M},{s}}^n \},$.
% \begin{equation}
%     {\chi}_{s}^n = \{\chi_{\mathsf{C},{s}}^n, \chi_{\mathsf{S},{s}}^n, \chi_{\mathsf{M},{s}}^n \},
% \end{equation}
Assume, $x_{m^\mathfrak{z}_s,n} \in \{0,1\},$ is a binary variable describing whether the VNF $m^\mathfrak{z}_s$ in layer  $\mathfrak{z} \in \{c, d\}$ in slice $s$ is utilizing the data center $n$ or not \cite{motalleb2019joint}.

%\textcolor{red}{
In the following, we will introduce an AI/ML method to optimize this system model. 
In addition, in this study, we consider a potential adversarial attack on our AI/ML approach which is a black-box attack (i.e., no knowledge).
 Attackers lack knowledge of our model and employ a weak adversary method to manipulate state and reward during agent interactions.
 Therefore, we require a secured technique to defend our system from these threats and vulnerabilities.%}
\section{DRL-based Energy-efficient Service Admission Control}~\label{drlsac}
In this section, firstly the problem formulation is obtained. The proposed method is examined on two different time scales. An estimation of a pre-defined number of VNFs is achieved on a large time scale. Next, the dynamic admission control and power minimization are solved on a small time frame.
%\vspace{-.2cm}
\subsection{Problem Statement}\label{probfor}
%\vspace{-.06cm}
%In this section, the problem formulation is described.
Assume the priority of the service $s$ is indicated with $p_s$. 
Moreover, for each data center $n$ that is hosting the VNF  $m^\mathfrak{z}_s$ in layer  $\mathfrak{z} \in \{c, d\}$ in slice $s$, the power consumption of the baseband processing can be represented as $\phi_{m^\mathfrak{z}_s,n}$.
Therefore, the total power consumption of all running data centers that are hosting the VNFs can be expressed as $ \textstyle \phi_{tot} = \sum_{n=1}^{N} \sum_{s=1}^{S} \sum_{m^\mathfrak{z}_s=1}^{M^\mathfrak{z}_s}  x_{m^\mathfrak{z}_s,n}\phi_{m^\mathfrak{z}_s,n}, \;\; \mathfrak{z} \in \{c, d\}.$
%\\
% &+ \sum_{n=1}^{N} z_n \phi_{n}, \;\; \mathfrak{z} \in \{c, d\}.$
% \begin{equation}
%     \textstyle \phi_{tot} = \sum_{n=1}^{N} \sum_{s=1}^{S} \sum_{m^\mathfrak{z}_s=1}^{M^\mathfrak{z}_s}  x_{m^\mathfrak{z}_s,n}\phi_{m^\mathfrak{z}_s,n}, \;\; \mathfrak{z} \in \{c, d\}.%\\
% % &+ \sum_{n=1}^{N} z_n \phi_{n}, \;\; \mathfrak{z} \in \{c, d\}.
% \end{equation}
% Where, $z_n$ is shown that whether the $n^{th}$ data center is turned on or not and $\phi_n$ is a static cost when a  data center is active.
% \begin{equation}
%   z_n =
%     \begin{cases}
%       1 & \sum_{s=1}^{S} \sum_{m_s=1}^{M_s}  x_{m_s,n} \geq 1 \\
%       0 & \text{otherwise}
%     \end{cases}       
% \end{equation}  
As a result, the cost function for the placement of VNFs on the data centers is formulated as follows:
%\vspace{-.2cm}
\begin{equation}\label{eqpsi}
\textstyle  \varphi_{tot} = \phi_{tot} - \kappa \sum_{n=1}^{N}\sum_{m_s=1}^{M_s} p_s x_{m_s,n}
\end{equation}
Where $\kappa$ is a design factor between the first term of \eqref{eqpsi}, representing the whole power of the resources, and the second term, is the total number of slices admitted with resources.
We aim to minimize the power and maximize the admitted rate with the presence of constraints as follows:
\vspace{-.2cm}
\begin{subequations}\label{q1}
\begin{alignat}{4}
\min\limits_{\boldsymbol{X}, \boldsymbol{M}}   \quad & {\varphi_{tot}} \\
\text{subject to} \quad  
 &\textstyle \sum_{s=1}^{S} \sum_{m_s=1}^{M_s}  x_{m_s,n} \bar{\psi}_{{\mathsf{C}},s}^{\mathfrak{z},tot}  \leq   \chi_{\mathsf{C},{s}}^n \;\; \forall n, \label{c1}\\
 &\textstyle \sum_{s=1}^{S} \sum_{m_s=1}^{M_s}  x_{m_s,n} \bar{\psi}_{{\mathsf{S}},s}^{\mathfrak{z},tot}  \leq   \chi_{\mathsf{S},{s}}^n  \;\;\forall n, \label{c2}\\
 &\textstyle \sum_{s=1}^{S} \sum_{m_s=1}^{M_s}  x_{m_s,n} \bar{\psi}_{{\mathsf{M}},s}^{\mathfrak{z},tot}  \leq   \chi_{\mathsf{M},{s}}^n  \;\;\forall n,  \label{c3}\\
 &\textstyle x_{m_s,n} \in \{0,1\}  \;\; \forall n, \forall s,  \forall m_s \label{c4}\\
 & \bar{T_s}^{tot} \leq T_{max}^s\label{c5}.
\end{alignat}
\end{subequations}
where,  $\mathfrak{z} \in \{c, d\}$, and the constraints \eqref{c1}, \eqref{c2}, and \eqref{c3} specify that VNFs hosted by data center $n$ cannot exceed the data center's total resources of CPU, memory, and the storage.
Moreover, \eqref{c4}, represents that the $x_{m_s,n}$ is a binary variable.
In addition, \eqref{c5}, indicates that the mean total delay of the system is less than the threshold.
Moreover, $\boldsymbol{X}$ is the matrix of the $x_{m^\mathfrak{z}_s,n}, \forall n, \forall m^\mathfrak{z}_s$ which defines the allocation of VNFs of slices to the resources of data centers.
Furthermore, $\boldsymbol{M}$ is the vector of $M_s,$  $\forall s$ that defines the number of pre-defined VNFs for each slice in the system.
\vspace{-.1cm}
\subsection{Proposed Method} \label{proposed}
In the following, we present our approach for addressing the problem outlined in \eqref{q1}, which necessitates solving it across two distinct time scales.
In the large time scale, we find the optimal number of pre-defined VNFs based on the mean arrival delay and the mean service time of the system at different times of network traffic. In the small time scale, we consider the problem of the dynamic service admission control based on the resource management of the VNFs in the system.
Due to the dynamic nature of the small time-scale problem, we are able to solve it using deep reinforcement learning (DRL).
Therefore, firstly, we can simplify the constraint \eqref{c5} and find the sub-optimal value for the number of the VNFs in each slice $s$. Afterward, we use the DRL technique to solve the problem of the admission control system.
\subsubsection{Estimation of The VNF Number}\label{lifecycle}
In this section, we want to find the optimal number of VNFs in the system for each slice in the large time scale based on the mean service time and mean arrival service rate.

In the lifecycle of the network slicing technique, we have four phases: Preparation, Commissioning, Operation and Decommissioning. The VNF numbers are estimated in the Preparation and Commissioning phase. However, in the Operation phase, it can be modified based on any change in the traffic of the system. However, with correct estimation, the power consumption of the system is reduced. 

We can simplify and relax the constraint \eqref{c5}. 
This constraint can be converted as $M_s \geq  \frac{\bar{\alpha}_s}{\bar{\mu}_s -2/\bar{T}_{\text{max}}} $.
% \begin{equation}
% M_s \geq  \frac{\bar{\alpha}_s}{\bar{\mu}_s -2/\bar{T}_{\text{max}}}    
% \end{equation}
Since we want to minimize the power consumption in the first term of the cost function of the problem \eqref{q1}, we consider the minimum value for the $M_s$, $\forall s$.
As a result, since the number of VNFs is the integer, we have $M_s = \lceil{\frac{\bar{\alpha}_s}{\bar{\mu}_s -2/\bar{T}_{\text{max}}}}\rceil$.
\subsubsection{Resource Management}
This section introduces a DRL-based network slicing resource management for dynamic service admission control and power minimization in the small time scale after solving the sub-optimal number of VNFs. This process is done in the Commissioning and Operation phase of network slice life cycle.

In the O-RAN architecture, the DRL method is carried out in the xApp in the near RT RIC.
The DRL approach combines deep neural networks (DNN) with reinforcement learning (RL).
We use a DRL method to solve this problem since we have a dynamic system.

All RL techniques represent a Markov decision-making process with $(\mathcal{S}, A, R, P, \gamma) $. Firstly, $A$ represents the action vector. $\mathcal{S}$ represents the state space matrix. Moreover, $R_t$ is the accumulated reward function and $r_t$ is a reward for taking action at time slot $t$. A probability of transfer is given by $ P (. | \mathcal{S}, a) $. Furthermore, The discount factor is defined as $ \gamma \in (0,1] $.
Moreover,  the $ \Pi (. | \mathcal{S}) $ is the policy that maps the state to the distribution of actions. In addition, the value-state function for state $\mathcal{s}$ under the policy $ \Pi (. | \mathcal{S}) $ with $ V^{\Pi} (\mathcal{s}) $ denotes the expected return value in state $\mathcal{s}$ under policy $ \Pi (. | \mathcal{S}) ) $.
Finally, The value of performing operation $a$ in state $\mathcal{s}$ under the $ \Pi (. | \mathcal{S}) $ policy is shown as $ Q ^ {\Pi} (\mathcal{s}, a) $.
% \begin{itemize}
% \item $A$ represents the action vector.
% \item $\mathcal{S}$ represents the state space matrix. 
% \item $R_t$ is the accumulated reward function and $r_t$ is a reward for taking action at time slot $t$. 
% \item A probability of transfer is given by $ P (. | \mathcal{S}, a) $.
% \item The discount factor is defined as $ \gamma \in (0,1] $.
% \item Moreover,  the $ \Pi (. | \mathcal{S}) $ is the policy that maps the state to the distribution of actions. 
% \item The value-state function for state $\mathcal{s}$ under the policy $ \Pi (. | \mathcal{S}) $ with $ V^{\Pi} (\mathcal{s}) $ denotes the expected return value in state $\mathcal{s}$ under policy $ \Pi (. | \mathcal{S}) ) $.
% \item The value of performing operation $a$ in state $\mathcal{s}$ under the $ \Pi (. | \mathcal{S}) $ policy is shown as $ Q ^ {\Pi} (\mathcal{s}, a) $.
% \end{itemize}

In the RL method, the aim is to maximize the total reward specified as $R_t =\sum_{k=0}^{\infty}\gamma^k r_{t+k}$. 
% \begin{equation}
% R_t = r_t + \gamma r_{t+1} + \gamma^2 r_{t+2} + ... = \sum_{k=0}^{\infty}\gamma^k r_{t+k}
% \end{equation}
%\vspace{-.02cm}
\subsubsection{Enviroment}\label{env}
This section introduces the Markov decision process (MDP) to describe an agent and environment based on the system model in Section \ref{system}.
\begin{itemize}
\item State: 
The state is the position of the agents at a specific time.
Assume in time step $t$, we have $r_t^s$ request from slice $s$.
In this problem, the state in time step $t$ is $\mathbb{S}_t = \{\boldsymbol{\chi}_t, \boldsymbol{R}_t\}$.
Where $\boldsymbol{\chi}_t \in \mathbb{C}^{N \times 3} $ is the 2D vector of remaining CPU, storage and memory for all data centers in time step $t$.
Furthermore, $\boldsymbol{R}_t \in \mathbb{C}^{S}$ is the 1D vector of service requests in time step $t$.
\item {Action}:
The action in each time step $t$ is represented by $\mathbb{A}_t = \{\boldsymbol{X}_t\}$, where $\boldsymbol{X}_t \in \mathbb{C}^{S\times N}$ is the 2D vector indicating whether the VNF of slice $s$ is assigned to the data center $n$ or not. 
\item Reward:
The aim is to maximize the admission rate and to minimize the number of activated data centers. 
The reward in each time step $t$ is defined by $\mathbb{R}_t$,
%\vspace{-.2cm}
\begin{equation}
    \mathbb{R}_t =     \begin{cases}
     \varphi_{tot, t}, & \chi^n_{i,s} \geq 0 \;\;\forall n, i \in \{C,S,M\}  \\
     -M  & \text{otherwise}
    \end{cases}
\end{equation}
where $\varphi_{tot, t}$ is the cost function of the system in each time step $t$. Moreover, $M$ is a large integer number.
\end{itemize}
%\vspace{-.1cm}
As mentioned in Section \ref{probfor}, in the problem \ref{q1}, the action is a discrete binary vector, and the state is continuous. Hence, we use the proximal policy optimization (PPO) method to solve this problem.
\subsubsection{PPO method}
A Temporal Difference (TD) representation of the Policy gradient can be seen in the Actor-Critic model.
In the actor-critic model, the system has two networks: the actor and the critic.
Based on the actor's decision, the action is taken. The actor learns by applying the policy gradient method. The actor receives feedback on the correctness of the action from the critic network.
The critic analyzes the actor using the value function.
A PPO is a method based on actor-critic analysis \cite{dynamicMelik}. 

The PPO algorithm is a policy gradient algorithm that balances simplicity, complexity, and tuning. By updating each step, it maintains a moderately low deviation from the previous policy. PPO is a reliable and efficient version of the trust region policy optimization (TRPO) algorithm that employs first-order optimization. Consequently, PPO combines actor-critic and TRPO concepts.
It is critical to note that the TRPO technique ensures that the updated policy is not too different from the old policy. Hence, the updated policy is within the trust region of the old policy. 
The objective function of TRPO can be formulated as follows.
%\vspace{-.3cm}
\begin{subequations}\label{problem}
\begin{alignat}{4}
\max_{\theta}& \quad \hat{\mathbb{E}}_t[\frac{\pi_{\theta}(a_t|\mathcal{s}_t)}{\pi_{\theta_{old}}(a_t|\mathcal{s}_t)}\hat{A_t}]\\
\text{subject to} \quad  &  \hat{\mathbb{E}}_t[\text{KL}[\pi_{\theta_{old}}(.|\mathcal{s}_t),\pi_{\theta}(a_t|\mathcal{s}_t)]] \leq  \delta,
 \label{c11} 
\end{alignat}
\label{constraints}
\end{subequations}
where $\pi_{\theta}$ is a stochastic policy and $\pi_{\theta_{old}}$ is the policy vector before updating. Moreover, $\hat{\mathbb{E}}_t[.]$ is the average of several samples and $\hat{A_t}$ is the advantage function estimator in the time of $t$.
In the TRPO method, in order to enable the trust region for optimization, KL divergence constraints must be met.
By modifying the clipped substitute objective function, PPO applies the policy constraint.
Assume $r_t(\theta) = \frac{\pi_{\theta}(a_t|\mathcal{s}_t)}{\pi_{\theta_{old}}(a_t|\mathcal{s}_t)}$.
In the PPO method the main objective function is $L^{CLIP}(\theta) = \hat{\mathbb{E}}_t[\min(r_t(\theta)\hat{A_t}, \text{clip}(r_t(\theta),1-\epsilon, 1+\epsilon)\hat{A_t})]$.
% \vspace{-.2cm}
% \begin{equation}
% L^{CLIP}(\theta) = \hat{\mathbb{E}}_t[\min(r_t(\theta)\hat{A_t}, \text{clip}(r_t(\theta),1-\epsilon, 1+\epsilon)\hat{A_t})]
% \end{equation}
where $1-\epsilon$ and $1+\epsilon$ are the lower and upper clipping ranges for state action $(\mathcal{s}, a)$. Moreover, the $r_t(\theta)$ is clipped to the lower and upper bound if it is out of this range \cite{wang2019trust}. 
\vspace{-0.1cm}
\section{Secure MTD Service Admission Control}\label{secur}
%\vspace{-0.1cm}
This section aims to secure the proposed energy-efficient service admission control against adversarial attacks.
%attack's dynamic service admission control and VNF placement in the O-RAN architecture. 
Firstly, we introduce the feasible attack model considered in this study. Then, we present the MTD strategy proposed to secure the system.
As depicted in Fig.~\ref{fig:c15}, the DRL-based service admission control service can be deployed as a xApp in the Near-RT RIC.
%The O-RAN Alliance introduce the xApps to perform the DRL method is performed in the near-RT RIC. 
To apply the MTD technique, the system requires different trained models. Each model is learned and deployed in a specific xApp. Therefore, we have different xApps consisting of DRL methods with similar performance.
\vspace{-.4cm}
\subsection{Adversarial Attack Model}\label{attackmodel}
We describe a feasible malicious attack on the proposed DRL method.
%for the architecture described previously.
There are three types of attacking the ML system based on the attacker's knowledge of the targeted model (i.e., model's parameters and architecture) and training data. The adversarial attack is considered white-box, gray-box, or black-box when the attacker obtains full, partial, or no knowledge, respectively~\cite{aisecme}.  
%access to every parameter and the whole of the system. The second one is the black box attack that the attacker can not access to parameters of the system, and the third one is the gray box attack that the attacker has access to some parameters of the system \cite{wang2021delving}.
Here, we assume that the adversary is targeting the PPO models under a black-box setting, i.e., no knowledge on the targeted model. 
%The black-box attacks in this paper are considered to be the target DRL agents.
To attack the system, we apply a weak adversary attack based on \cite{wu2021reinforcement}.
Assume that attackers want to attack the system at time $t$.
The attackers generate an arbitrary state $\hat{s}_t$ and the corresponding reward function $\hat{r}(\hat{s}_t,.)$.
 After the agent determines the altered state $\hat{s}_t$, it carries out the action $a_t$ and observes $\hat{r}(\hat{s}_t,a_t)$, instead of $r(s_t,a_t)$.
%  As a result, we assume that a perturbation occurs in the state of the system in each time step, which is the remaining resources and the service arrival rate of the services.
%  Simulations were conducted by altering the service arrival rates and converting them to a uniform random variable between zero and the service arrival rate. Based on the weak adversary attack in \cite{wu2021reinforcement}, we blocked part of the service arrival rates.
\vspace{-.1cm}
\subsection{Moving Target Defense Strategy}
MTD, an emerging security strategy, continuously alters system configurations, making attacks challenging due to increased uncertainty and complexity. This method lowers attack success rates by reducing attacker knowledge and effectiveness. MTD enhances defense by adding ambiguity and offering multiple configurations
 \cite{sengupta2019mtdeep}.
 
We deploy four diverse PPO models with varied configurations into different xApps in this scenario.  %\textcolor{red}{
These four methods give us almost similar results but with different configurations in terms of the number of neural network layers, batch size, discount factor, learning rates, among others.%}
It is critical to note that attackers have a set of attacks that are designed to attack the configurations of the defender.
%\textcolor{red}{
In the process of training, the attacker randomly chooses one of the xApps that contains one of these PPO models and attacks it.%}
Hence, once these four models have been trained, one random model is chosen from the four to run on each input and return the output generated by that model.
%\textcolor{red}{
As attacks are directed at one of the models, attackers have less impact on the system since they do not know which model the system selects at a given time. Therefore, the probability of an adversarial attack against our system is diminishing by randomly choosing one of these models that can be the un-attacked model.%}
\vspace{-.2cm}
\section{Numerical results} \label{numerical}
This section presents numerical results for the main problem. Considering the similarity of packets between O-CU and O-DU, their requirements are equivalent. Only minor headers in O-CU packets are removed in O-DU, having negligible impact on processing. Therefore, we assume O-CU and O-DU share the same processors (VNFs).
In these figures, we consider two data centers, each equipped with a CPU boasting 32 cores, 50GB memory, and 5TB storage.

Assume there are two service requests. Each service is assigned to a specific slice.
Each slice contains $M_s$, pre-defined similar VNFs that are obtained from the large time scale.
 For the first service, each request needs 2 cores, 7GB memory, and 30GB storage in O-CU and O-DU. The second service requires 3 cores, 5GB memory, and 50GB storage per request in O-CU and O-DU.

To assess the performance of the proposed solutions, we illustrate five different scenarios. The first scenario is the exhaustive search. 
The system works with the PPO model in the second scenario without attack. The third scenario involves an adversarial attack on the system without protection. In the fourth scenario, the protected MTD system is under attack. The fifth scenario considers the baseline method, in which random allocation is assumed.
The training process involves learning four PPO models with different parameters (different batch sizes, discount factor, learning rate, the number of steps to run for each environment, etc. ). The models have similar performance. At each time slot, one of these models is chosen randomly to protect the system from attack.

Fig. 2 displays the mean reward over time slots for a system without service admission control attacks. This system features 12 service arrival rates per time slot for two distinct services with varying QoS, demonstrating PPO convergence. 
% \begin{figure*}%[htb]
%     \centering
%     \subfigure[\footnotesize{Mean Reward vs. Time Slots}]{\label{fig:f3}\includegraphics[width=0.3\textwidth]{reward.eps}}
%     \centering
%     \subfigure[\footnotesize{Service Admission Rate vs. Service Arrival Rate}]{\label{fig:f1}\includegraphics[width=0.3\textwidth]{bar1_final.eps}}
%     \centering
%     \captionsetup{justification=centering}
%     \subfigure[Power Consumption vs. Service Arrival Rate]{\label{fig:f2}\includegraphics[width=0.3\textwidth]{bar6.eps}}
%     \captionsetup{justification=centering}
%     %\caption{Performance results of secured dynamic service admission control and power consumption.}
%     \label{fig:all}
% \end{figure*}
\begin{figure*}[!htb]
	\centerline{%
		\begin{tabular}{c@{}c@{}c}			
			\includegraphics[scale = 0.35]{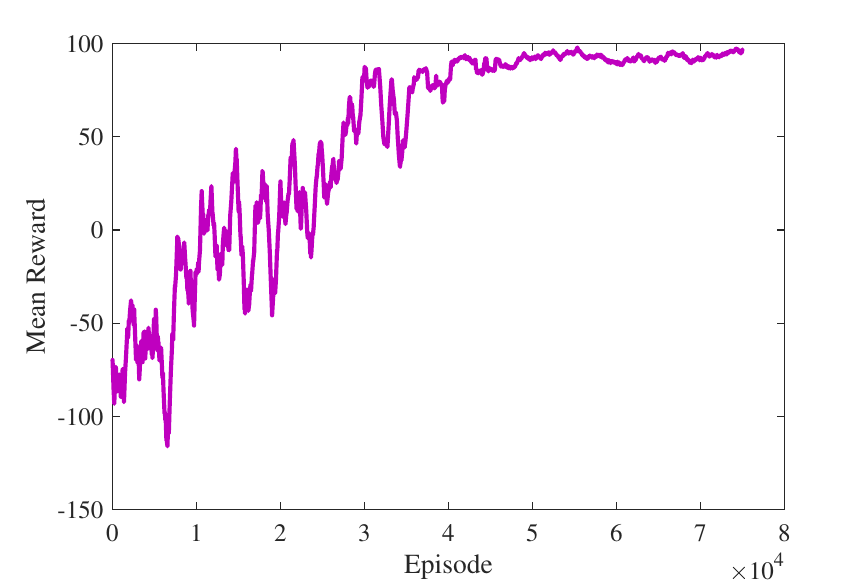} &
			\includegraphics[scale = 0.35]{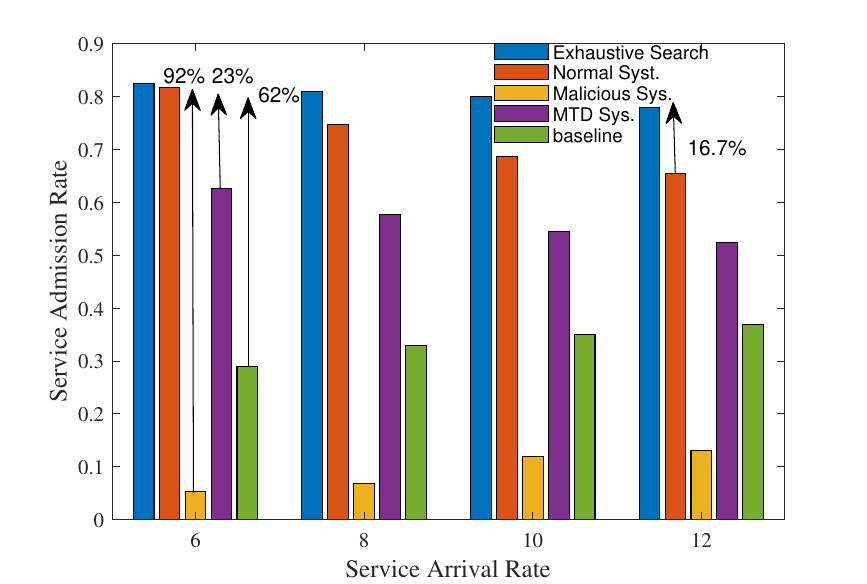} &
			\includegraphics[scale = 0.35]{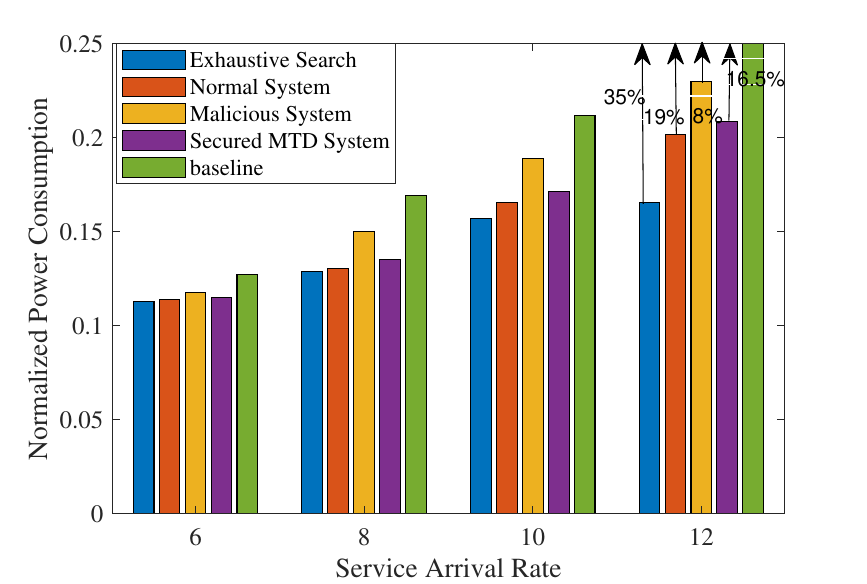} \\
			\scriptsize Fig. 2~Mean Reward vs. Time Slots. & \scriptsize  Fig. 3~Service Admission Rate vs. Service Arrival Rate. & \scriptsize Fig. 4~Power Consumption vs. Service Arrival Rat.e 
	\end{tabular}}
	%\caption{Global caption.}
	\label{label}
 \vspace{-0.71cm}
\end{figure*}
In Fig. 3, the service admission rate is depicted for different service arrival rates for the five scenarios.
In this simulation, we assume that the design factor $\kappa$ is large enough that the cost function is affected just by the admission control system, and the power consumption is not considered here.
The figure shows a dynamic service arrival in a system. Each time slot has a $30\%$ service departure rate. 
This figure indicates that over $80\%$ of admissions were recorded whenever we had the average of six service arrival rates for each service in each time slot in the system without any attack.
As service arrival rates rise, admission rates fall due to increased packet arrivals and traffic, leading to reduced admission rates at the system's fixed capacity 
The system's performance decreases by at most 93\% when under attack, showing the considerable impact adversarial attacks can have on unprotected ML models. The MTD-protected system has significantly improved the system's robustness, yielding a $70\%$ increase in service admission rate compared to the attacked system.
Moreover, the system's performance increases $62\%$ compared to the baseline, which is the random allocation.
The optimality of the PPO model decreases to 16.7\% as the service arrival rate increases to 12.
%Compared to the normal system, the protected system has a 23\% loss. Consequently, it has increased performance by 70\% from the attacked system.

% \begin{figure}
%   \centering
%   \captionsetup{justification=centering}
%   %\includegraphics[width=\textwidth/2.2]{ocloud4.pdf}
%     \includegraphics[scale = .38]{reward.eps}
%     %\includegraphics[width=9.cm,height=6.2cm]{finalDraw1.pdf}
%     %\includegraphics[width=\textwidth]{finalDraw.pdf}
%   \caption{Mean Reward vs. Time Slots}
%   \label{fig:f3}
% \end{figure}
% \begin{figure}
%   \centering
%   \captionsetup{justification=centering}
%   %\includegraphics[width=\textwidth/2.2]{ocloud4.pdf}
%     \includegraphics[scale = .38]{fig1_final.eps}
%     %\includegraphics[width=9.cm,height=6.2cm]{finalDraw1.pdf}
%     %\includegraphics[width=\textwidth]{finalDraw.pdf}
%   \caption{Service Admission Rate vs. Service Arrival Rate}
%   \label{fig:f1}
% \end{figure}
% \begin{figure}
%   \centering
%   \captionsetup{justification=centering}
%   %\includegraphics[width=\textwidth/2.2]{ocloud4.pdf}
%     \includegraphics[scale = .38]{bar5.eps}
%     %\includegraphics[width=9.cm,height=6.2cm]{finalDraw1.pdf}
%     %\includegraphics[width=\textwidth]{finalDraw.pdf}
%   \caption{Mean Power Consumption vs. Service Arrival Rate}
%   \label{fig:f2}
% \end{figure}
In Fig. 4, we present the normalized power consumption across five scenarios, varying with service arrival rates. The parameter $\kappa$ relates power consumption to admission control, albeit power usage remains substantial relative to admission control. Nevertheless, admission control's impact is non-negligible.
In our simulation, baseband processing power $\phi_{m_s,n}$ is uniformly distributed within the range [100,200]. The figure highlights that higher service arrival rates correspond to increased system power consumption. 
The malicious system consumes more power due to the inverse power-reward relationship. Despite denied control requests from attacks, the attacker aims to maximize power, vital for DRL rewards.
The power of the normally trained system is decreased $19\%$ compared to the baseline system, which is the random allocation. 
The power of the PPO system is increased $11\%$ (based on the baseline method) after the system is attacked. Whenever we apply the MTD technique to an attacked system, the power is reduced by 8.5\%. Also, the power consumption of the PPO system increased 16\% (based on the baseline method) compared to the optimal method.
Assume the delay of the service in the system is $T_s = 1.07 
\mu sec$.
The mean service rate is considered to be $2$ Mbps and the mean arrival rate of the system is $1$ Mbps. The estimated number of VNFs is $8$. In Table \ref{table:1}, the extra power consumption ratio for $10$ and $12$ VNFs is obtained. If we apply $10$ VNFs instead of $8$ VNFs, the mean power increases by $21\%$. Considering $12$ VNFs instead of $8$ VNFs, the mean power increased $27\%$.
In the network slicing life cycle (as in \ref{lifecycle}), optimal VNF estimation happens during preparation and commissioning, the planning phase. However, adjustments are possible during operation, reducing excess power use caused by inaccurate initial estimates
\begin{table} %[H]
 \caption {un-estimated VNF numbers vs. the extra power consumption } \label{table:1a}
 \begin{center}
 \scalebox{0.6}{
  \begin{tabular}{l  l }
  \toprule
\textbf{Number of VNFs} & \textbf{Extra Power Consumption} \\ [0.4ex]
  \toprule\toprule
  12 & 27 \%\\
  10 & 21 \%  \\
 \toprule
 \end{tabular}}
 \vspace*{-3em}
 \end{center}
 \label{table:1}
 \end{table}
\vspace{-.2cm}
 \section{Conclusion} \label{conclud}
 \vspace{-.1cm}
In this academic study, we address two pivotal challenges within the O-RAN architecture: firstly, the determination of the optimal number of VNFs for each slice, and secondly, the establishment of secure AI/ML methodologies for the dynamic management of service admission control and power reduction within the O-RAN framework. We derive sub-optimal VNF quantities at a larger time scale and employ an actor-critic approach with the PPO algorithm at a smaller scale. Four PPO models are trained in distinct xApps within the near RT RIC. Security is bolstered using MTD, randomly selecting xApps with trained PPO models for added unpredictability. Numerical results demonstrate robust PPO performance in the absence of attacks, significantly improved by the MTD strategy in adversarial scenarios. 
%\textcolor{red}{
To enhance system security with MTD, we need to train multiple models, despite its limitations. Future MTD techniques should aim for a balance between robustness, performance, and cost.
  \vspace{-.15cm}
\section*{Acknowledgment}
This research work is partially supported by the Business Finland 6Bridge 6Core project under Grant No. 8410/31/2022, the Research Council of Finland (former Academy of Finland) IDEA-MILL project (Grant No. 352428), the Research Council of Finland (former Academy of Finland) 6G Flagship program (Grant No. 346208), and the European Union’s Horizon Europe research and innovation programme HORIZON-JU-SNS-2022 under the RIGOUROUS project (Grant No. 101095933) and the 6G-SANDBOX project (Grant No. 101096328). The paper reflects only the authors' views. The Commission disclaims responsibility for any use of the information it contains.
\vspace{-.2cm}
\bibliographystyle{IEEEtran}
\bibliography{ref}
\end{document}